\documentclass[conference,final]{IEEEtran}
\usepackage{cite}
\usepackage{amsmath,amssymb,amsfonts}
\usepackage{algorithmic}
\usepackage{graphicx}
\usepackage{textcomp}
\usepackage{xcolor}
\ifCLASSOPTIONcompsoc
\usepackage[caption=false,font=normalsize,labelfont=sf,textfont=sf]{subfig}
\else
\usepackage[caption=false,font=footnotesize]{subfig}
\fi

\usepackage{mathtools}
\usepackage{siunitx}
\usepackage{bm}
\usepackage[acronym]{glossaries-extra}
\usepackage{booktabs}
\usepackage{threeparttable}

\newcommand{\vt}[1]{\boldsymbol{\mathrm{#1}}}
\newcommand{\mt}[1]{\boldsymbol{\mathrm{#1}}}
\newcommand{\tr}[1]{\text{tr}\left(#1\right)}
\newcommand{\CN}[1]{\mathcal{CN}\left(#1\right)}
\DeclareMathOperator*{\diag}{diag}
\DeclareMathOperator*{\rank}{rank}
\DeclareMathOperator*{\argmax}{arg\,max}
\DeclarePairedDelimiter{\floor}{\lfloor}{\rfloor}
\DeclarePairedDelimiter{\abs}{\lvert}{\rvert}
\DeclarePairedDelimiter{\norm}{\lVert}{\rVert}

\DeclareSIUnit{\dBmpHz}{dBm/Hz}
\DeclareSIUnit{\dBm}{dBm}

\makeglossaries
\setabbreviationstyle[acronym]{long-short}
\newacronym{irs}{IRS}{intelligent reflecting surface}
\newacronym{bbu}{BBU}{base band unit}
\newacronym[firstplural=remote radio heads]{rrh}{RRH}{remote radio head}
\newacronym{psd}{PSD}{power spectral density}
\newacronym{los}{LoS}{line-of-sight}
\newacronym{sdr}{SDR}{semi-definite relaxation}
\newacronym{iid}{i.i.d.}{independent and identically distributed}
\newacronym{pdf}{PDF}{probability density function}
\newacronym{glrt}{GLRT}{generalized likelihood ratio test}
\newacronym{mimo}{MIMO}{multiple-input multiple-output}
\newacronym{snr}{SNR}{signal-to-noise ratio}
\newacronym{cdf}{CDF}{cumulative distribution function}
\newacronym{bs}{BS}{base station}
\newacronym{iot}{IoT}{Internet of Things}
\newacronym{ris}{RIS}{reconfigurable intelligent surface}
\newacronym{csi}{CSI}{channel state information}
\newacronym{sinr}{SINR}{signal-to-interference-plus-noise ratio}

\def\BibTeX{{\rm B\kern-.05em{\sc i\kern-.025em b}\kern-.08em
    T\kern-.1667em\lower.7ex\hbox{E}\kern-.125emX}}
\begin{document}
\setlength{\textheight}{680pt}
\setlength{\voffset}{-1pt}
\setlength{\textwidth}{520pt}
\setlength{\hoffset}{-2pt}

\bstctlcite{IEEEexample:BSTcontrol}
\title{IRS-Assisted Active Device Detection}
\author{Friedemann~Laue\IEEEauthorrefmark{1}\IEEEauthorrefmark{2}, Vahid~Jamali\IEEEauthorrefmark{1}, and Robert~Schober\IEEEauthorrefmark{1}\\
\IEEEauthorrefmark{1}\textit{Friedrich-Alexander University Erlangen-Nürnberg (FAU)},\IEEEauthorrefmark{2}\textit{Fraunhofer IIS}}

\maketitle

\begin{abstract}
  This paper studies intelligent reflecting surface (IRS) assisted active device detection. Since the locations of the devices are a priori unknown, optimal IRS beam alignment is not possible and a worst-case design for a given coverage area is developed. To this end, we propose a generalized likelihood ratio test (GLRT) detection scheme and an IRS phase-shift design that minimizes the worst-case probability of misdetection. In addition to the proposed optimization-based phase-shift design, we consider two alternative suboptimal designs based on closed-form expressions for the IRS phase shifts. Our performance analysis establishes the superiority of the optimization-based design, especially for large coverage areas. Furthermore, we investigate the impact of scatterers on the proposed line-of-sight based design using simulations.
\end{abstract}

\section{Introduction}

\Glspl{irs} have gained significant attention due to their capability of transforming the wireless channel into a programmable smart radio environment \cite{renzo2020smartradioenvironments}. An \gls{irs} comprises a large number of unit cells, which are configured to induce specific phase shifts to an impinging electromagnetic wave. Given an optimized phase-shift design for the unit cells, an \gls{irs} can significantly improve the performance of the communication system \cite{wu2019intelligentreflectingsurface}.
However, most works in the literature propose phase-shift designs for an active communication link and do not consider use cases where the activity and the location of a device are a priori unknown.
The detection of active devices is needed, e.g., in \gls{iot} networks, where sensors provide sporadic status reports, and in the initial access stage of cellular communication systems \cite{habib2017millimeterwavecell,yan2019compressiveinitialaccess}.
For such applications, since no communication link exists prior to the successful detection of the active devices, the phase-shift designs proposed in the literature are not applicable.
Consequently, alternative phase-shift designs are required that provide basic connectivity over an \gls{irs}-assisted channel at any time and regardless of the devices' locations.

To this end, this paper studies the detection of active devices in \gls{irs}-assisted communication systems. We assume a given coverage area, where devices sporadically access the \gls{bs} for data transfer by transmitting known synchronization signals, while the exact locations of the active devices are unknown. Furthermore, we assume that the direct link between the devices and the \gls{bs} is blocked and an \gls{irs} is deployed to achieve connectivity. In order to find suitable phase-shift designs for the \gls{irs}, we first derive a \gls{glrt} detector for the synchronization signals based on a physics-based model of the \gls{irs}-assisted end-to-end channel. Subsequently, we propose an optimized phase-shift design that minimizes the probability of misdetection for the devices' worst-case locations. Moreover, we study two heuristic analytical phase-shift designs and compare their performance to that of the optimized design for various channel conditions.

We note that phase-shift optimization has also been studied for the initial access in millimeter-wave communication networks \cite{aykin2020efficientbeamsweeping,yan2019compressiveinitialaccess,habib2017millimeterwavecell}. However, these results sythesize the beam pattern for a \gls{bs} for specific angles whereas the \gls{irs} phase-shift design has to create a reflection pattern that depends on both the incident and the reflection directions. Thus, the results for millimeter-wave communication networks are not directly applicable to \gls{irs}-assisted systems.

\textit{Notations}: $\tr{\mt{X}}$, $\mt{X}^T$, $\mt{X}^H$, and $\diag(\mt{X})$ denote the trace, transpose, conjuate transpose, and the vector comprising the elements on the main diagonal of matrix $\mt{X}$, respectively. A positive semidefinite matrix $\mt{X}$ is characterized by $\mt{X}\succeq 0$ and the identity matrix of size $N\times N$ is denoted by $\mt{I}_N$.
$[\vt{x}]_n$ refers to the $n$th element of vector $\vt{x}$. $\mathbb{R}$ and $\mathbb{C}$ denote the set of real and complex numbers, respectively. $\vt{x}\sim\mathcal{N}(\vt{a}, \mt{B})$ and $\vt{x}\sim\mathcal{CN}(\vt{a}, \mt{B})$ refer to normal and complex normal distributed random vectors $\vt{x}$ with mean vector $\vt{a}$ and covariance matrix $\mt{B}$, respectively.
The remainder of the division $a/b$ is denoted by $a\pmod{b}$ and the largest integer less or equal to $x$ is given by $\floor{x}$.

\section{System Model}\label{sec:systemmodel}

\begin{figure}
\centerline{\includegraphics[width=\columnwidth]{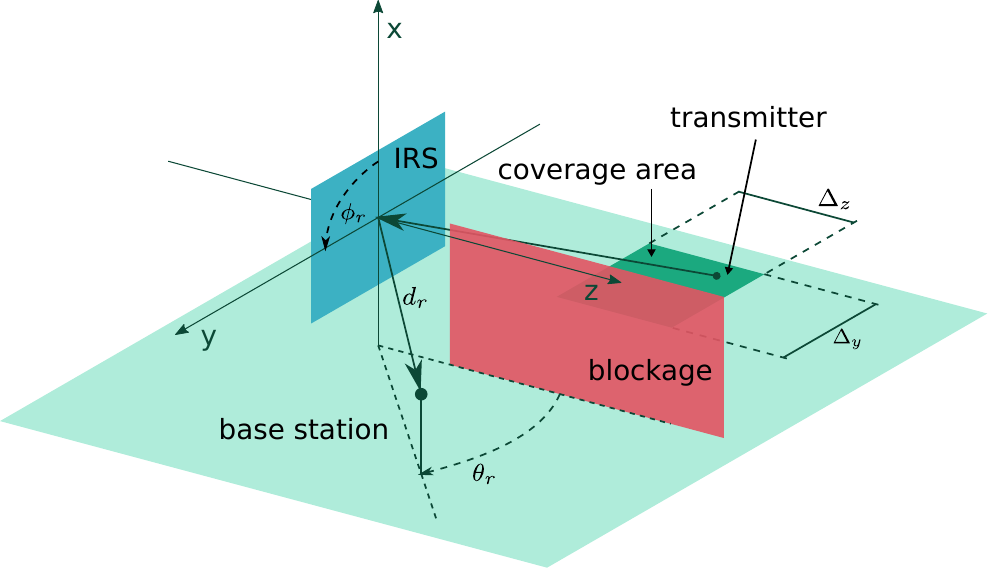}}
\caption{Schematic illustration of the considered IRS-assisted communication setup for active device detection.}
\label{fig:scene}
\end{figure}

Fig.~\ref{fig:scene} illustrates the considered communication scenario consisting of a \gls{bs}, an \gls{irs}, and a coverage area where the devices of interest are located. The blockage prevents a direct link between the devices and the \gls{bs}. Therefore, the \gls{irs} is deployed to reflect waves originating from the coverage area towards the \gls{bs}. The center of the \gls{irs} is the origin of two coordinate systems. In particular, for ease of presentation, we use a standard Cartesian coordinate system $(x, y, z)$ to characterize the coverage area and the \gls{irs} unit-cell locations, and a spherical coordinate system $(d, \theta, \phi)$ to characterize the location of the \gls{bs} and the incident and reflection angles on the \gls{irs}, see Fig.~\ref{fig:scene}.

\subsection{Coverage Area and Devices}

Two parameter sets define the rectangular coverage area in the $yz$-plane: the coordinates $(c_x, c_y, c_z)$ of the center of the area and dimensions $(\Delta_y, \Delta_z)$ along the $y$ and $z$ axes. In general, we use the coordinates $(q_x, q_y, q_z)$ to refer to a specific location $q$ in the coverage area, where $q_x=c_x$, $q_y\in[c_y - \Delta_y/2, c_y + \Delta_y/2]$, and $q_z\in[c_z - \Delta_z/2, c_z + \Delta_z/2]$.
Most of the time, a device in the coverage area is inactive and not synchronized to the \gls{bs}. When new data arrives in the transmit buffer of a device, it becomes active and attempts to access the \gls{bs} by transmitting a predefined synchronization signal $\vt{x}\in\mathbb{C}^{S}$  comprising $S$ symbols. We define a constant transmit power $P_x$ for each symbol and assume only one device is active at a time.

\subsection{Intelligent Reflecting Surface}

The \gls{irs} is located in the $xy$-plane and comprises $U=U_xU_y$ unit cells, where $U_x$ ($U_y$) is the number of unit cells along the $x$-axis ($y$-axis). The unit-cell spacing in $x$-direction ($y$-direction) is denoted by $d_x$ ($d_y$). We define set $\mathcal{U}=\{0, 1, \dots, U-1\}$ and index each unit cell by $u\in\mathcal{U}$ or by the two-dimensional index
\begin{align}
  u_x &= u\pmod{U_x} - U_x/2 + 1\\
  u_y &= \floor{u/U_x} - U_y/2 + 1,
\end{align}
where we assume that $U_x$ and $U_y$ are even numbers. The coordinates of the $(u_x, u_y)$th unit cell are given by vector $\vt{c}_{u_x,u_y} = \begin{bmatrix}d_x u_x & d_y u_y & 0\end{bmatrix}^T$. The phase shift of the $u$th unit cell $\varphi_u$ determines the $u$th element of phase-shift vector $\vt{w}\in\mathbb{C}^U$, i.e., $[\vt{w}]_u=e^{j\varphi_u}$.

Following the model in \cite{najafi2020physicsbasedmodeling}, we assume that both the transmitter (device) and the receiver (\gls{bs}) are located in the far field of the \gls{irs}.
Let $\vt{\Psi}_t^{(q)} = \begin{bmatrix}\theta_t^{(q)} & \phi_t^{(q)}\end{bmatrix}^T$ and $\vt{\Psi}_r = \begin{bmatrix}\theta_r & \phi_r\end{bmatrix}^T$ denote the direction from the \gls{irs} to a transmitter at location $q$ and to the \gls{bs}, respectively. Then, the \gls{irs} response function \cite{najafi2020physicsbasedmodeling}
\begin{equation}\label{eq:irsresponse}
  g(\vt{\Psi}_t^{(q)}, \vt{\Psi}_r) = \upsilon(\vt{\Psi}_t^{(q)}, \vt{\Psi}_r) \vt{a}^H(\vt{\Psi}_t^{(q)}, \vt{\Psi}_r) \vt{w},
\end{equation}
characterizes the reflected wave observed in direction $\vt{\Psi}_r$ caused by an impinging plane wave from direction $\vt{\Psi}_t^{(q)}$.
The elements of the steering vector $\vt{a}(\vt{\Psi}_t^{(q)}, \vt{\Psi}_r)$ in \eqref{eq:irsresponse} are given by
\begin{equation}
  \left[\vt{a}(\vt{\Psi}_t^{(q)}, \vt{\Psi}_r)\right]_{u_x,u_y} =
    e^{
      -j\left(\vt{k}(\vt{\Psi}_t^{(q)}) + \vt{k}(\vt{\Psi}_r)\right)^T
      \vt{c}_{u_x,u_y}
    }
\end{equation}
and
\begin{equation}
  \vt{k}(\vt{\Psi}) = \frac{2\pi}{\lambda}
    \begin{bmatrix}
      \sin(\theta)\cos(\phi) & \sin(\theta)\sin(\phi) & \cos(\theta)
    \end{bmatrix}^T,
\end{equation}
where $\lambda$ denotes the wavelength and $\vt{\Psi}$ refers to an incident direction $\vt{\Psi}_t^{(q)}$ or reflection direction $\vt{\Psi}_r$. Furthermore, $\upsilon(\vt{\Psi}_t^{(q)}, \vt{\Psi}_r)$ in \eqref{eq:irsresponse} denotes the unit cell factor and is specified in \cite[Section II.D]{najafi2020physicsbasedmodeling}.

\subsection{Channel Model}

We consider single-antenna devices and a multiple-antenna \gls{bs} with $M$ antenna elements. We assume \gls{irs} and \gls{bs} are deployed at sufficient height such that their \gls{los} is much stronger than any scattered links. However, since the devices are at low heights, several scatterers may contribute to the device-\gls{irs} channel. Hence, the end-to-end channel from location $q$ in the coverage area to the \gls{bs} is modelled as
\begin{equation}\label{eq:vhp}
  \vt{h}_q = \vt{h}_r \sum_{l=0}^{L-1} g(\vt{\Psi}_{t,l}^{(q)}, \vt{\Psi}_r)
    h_{t,l}^{(q)},
\end{equation}
where $\vt{h}_r=h_r\vt{b}$ denotes the \gls{irs}-\gls{bs} channel with $\vt{b}\in\mathbb{C}$ being the \gls{bs} steering vector. The $m$th phase shift $\varphi_m$ of $[\vt{b}]_m = e^{j\varphi_m}$ is relative to the reference phase $\varphi_r=2\pi d_r / \lambda$ of the \gls{irs}-\gls{bs} free-space channel
\begin{equation}\label{eq:hr}
  h_r = \frac{\lambda}{4\pi d_r}e^{j\varphi_r},
\end{equation}
where $d_r$ denotes the \gls{irs}-\gls{bs} distance. Moreover, $L$ in \eqref{eq:vhp} denotes the number of paths in the device-\gls{irs} link and $h_{t,l}^{(q)}$ denotes the channel coefficient of the $l$th path, where $l\in\{0,1,\dots,L-1\}$. The \gls{los} channel coefficient, $h_{t,0}^{(q)}$, is deterministic, whereas the non-\gls{los} channel coefficients are modelled as Rayleigh fading, i.e.,
\begin{align}
  h_{t,l}^{(q)}& = \frac{\lambda}{4\pi d_t^{(q)}}e^{j\varphi_t^{(q)}}, && l=0\label{eq:ht}\\
  h_{t,l}^{(q)}& \sim \mathcal{CN}\left(0, P_{l,\mathrm{NLoS}}^{(q)}\right), && l\neq 0.\label{eq:htl}
\end{align}
In \eqref{eq:ht}, $\varphi_t^{(q)}=2\pi d_t^{(q)} / \lambda$, where $d_t^{(q)}$ denotes the distance between device location $q$ and the \gls{irs}. In \eqref{eq:htl}, $P_{l,\mathrm{NLoS}}^{(q)}$ denotes the average power of the $l$th scattered path.
Furthermore, the incident directions $\vt{\Psi}_{t,l}^{(q)}$ for $l\neq 0$ of the scattered waves originating from location $q$ are modelled as random variables that are specified by a given probability distribution \cite{bjornson2020rayleighfadingmodeling}.

\section{Active Device Detection}\label{sec:detectiondesign}

For the design of the detection scheme at the \gls{bs}, we assume that only few scatterers exist in the device-\gls{irs} link and that the \gls{los} path contributes most power to the received signal. This assumption is justified for systems operating in the millimeter-wave frequency bands \cite{rappaport2013broadbandmillimeterwave}.
Thus, for tractability, we design the system based on the \gls{los} link only. However, in Section \ref{sec:results}, we will investigate the impact of scatterers in the device-\gls{irs} link on the proposed design.
Assuming only the \gls{los} link exists, the received symbols that originate from a device at location $q$ are given by
\begin{equation}\label{eq:yps}
  \mt{Y}_q = h_q \vt{b} \vt{x}^T + \mt{Z},
\end{equation}
where the element in the $m$th row and $s$th column of $\mt{Y}_q\in\mathbb{C}^{M\times S}$ denotes the symbol received at the $m$th \gls{bs} antenna and in the $s$th symbol interval, $s\in\{0,1,\dots,S-1\}$.
Moreover, the end-to-end channel coefficient is given by $h_q = h_r g(\vt{\Psi}_{t,0}^{(q)}, \vt{\Psi}_r) h_{t,0}^{(q)}$ and the elements of $\mt{Z}\in\mathbb{C}^{M\times S}$ are mutually independent complex normal random variables with zero mean and variance $\sigma_n^2$, denoting additive white Gaussian noise.
We assume $\vt{b}$ is given and fixed because its values only depend on the geometry of the \gls{bs} antenna. Therefore, we adopt a matched filter $\vt{v}^H = \vt{b}^H/\sqrt{M}$ at the \gls{bs} and obtain the filtered signal
\begin{equation}\label{eq:linearmodel}
  \vt{y}_q^T = \vt{v}^H \mt{Y}_q = \sqrt{M} h_q \vt{x}^T + \vt{z}^T
  = \vt{s}_q^T e_q + \vt{z}^T,
\end{equation}
where $\vt{z}^T = \vt{v}^H\mt{Z} \sim\CN{0, \sigma_n^2\mt{I}_{S}}$, $\vt{s}_q^T = \vt{x}^T \sqrt{M} \abs*{h_q}$, and $e_q = \exp(j\arg(h_q))$.
Here, $\arg(h_q)$ denotes the phase of $h_q$.
Although we can determine $e_q$ using \eqref{eq:irsresponse}, \eqref{eq:hr}, and \eqref{eq:ht}, its value depends on the exact knowledge of distances $d_r$ and $d_t^{(q)}$. In practice, these distances have to be obtained from measurements with a limited accuracy. Unfortunately, even small deviations (on the order of wavelengths) from the exact values may result in large variations of $\arg(h_q)$ and consequently of $e_q$. As a result, we model $e_q$ as a deterministic, but unknown variable. On the other hand, the impact of estimation errors for $d_r$ and $d_t^{(q)}$ on $\abs*{h_q}$ and equivalently $\vt{s}_q$ is less significant.
Hence, we assume that $\abs*{h_q}$ and $\vt{s}_q$ are known.

In general, for active-device detection, two hypotheses are defined and the detector decides for either of them given the observation \cite{kay1998fundamentalsstatisticalsignal}. For the problem at hand, the hypotheses are:
\begin{align}
  \mathcal{H}_0:&\ \text{device inactive}& \Rightarrow \vt{y}_q&= \vt{z}\\
  \mathcal{H}_1:&\ \text{device active}& \Rightarrow \vt{y}_q&= \vt{s}_q e_q + \vt{z}
\end{align}
For derivation of the detection scheme, we apply the \gls{glrt} concept, which replaces the unknown variable $e_p$ by its maximum likelihood estimate. Denote $f(\vt{y}_q; e_q, \mathcal{H}_1)$ as the \gls{pdf} of $\vt{y}_q$ under $\mathcal{H}_1$ with parameter $e_q$ and $f(\vt{y}_q; \mathcal{H}_0)$ as the \gls{pdf} of $\vt{y}_q$ under $\mathcal{H}_0$.
Then, the generalized likelihood ratio for considered scenario is given by
\begin{equation}\label{eq:likelihoodratio}
    L_G(\vt{y}_q) = \frac{\max_{e_q} f(\vt{y}_q; e_q, \mathcal{H}_1)}
      {f(\vt{y}_q; \mathcal{H}_0)}
\end{equation}
and the detector decides for $\mathcal{H}_1$ when \eqref{eq:likelihoodratio} is larger than a detection threshold $t^{\prime}$.
For $S>1$, it can be shown \cite[Chapter 7]{kay1998fundamentalsstatisticalsignal} that $L_G(\vt{y}_q) > t^{\prime}$ is equivalent to
\begin{equation}\label{eq:metric}
  T(\vt{y}_q) = 2\ln L_G(\vt{y}_q)
    = 2\frac{
      \abs*{\vt{s}_q^H \vt{y}_q}^2
      }{
        \sigma_n^2 \norm*{\vt{s}_q}^2
      } > 2\ln t^{\prime} = t.
\end{equation}
The distribution of $T(\vt{y}_q)$ under both hypotheses is
\begin{align}
  \mathcal{H}_0&:\quad T(\vt{y}_q) \sim \chi_{2}^2(0)\label{eq:ph0}\\
  \mathcal{H}_1&:\quad T(\vt{y}_q) \sim \chi_{2}^2\left(\gamma_q\right),\label{eq:ph1}
\end{align}
where $\chi_2^2(\gamma_q)$ denotes a $\chi^2$ distribution with $2$ degrees of freedom and non-centrality parameter
\begin{equation}\label{eq:gamma_q}
  \gamma_q = 2 S M \abs*{h_q}^2 \frac{P_x}{\sigma_n^2}.
\end{equation}
One can directly obtain the probability of false alarm $\Gamma_\mathrm{FA} = \mathrm{Pr}\{T(\vt{y}_q)>t|\mathcal{H}_0\}$ and the probability of misdetection $\Gamma_\mathrm{MD}^{(q)} = \mathrm{Pr}\{T(\vt{y}_q)<t|\mathcal{H}_1\}$ from \eqref{eq:ph0} and \eqref{eq:ph1} as \cite[Chapter 13.4]{kay1998fundamentalsstatisticalsignal}
\begin{align}
  \Gamma_\mathrm{FA}& =1-F_{\chi_{2}^2(0)}(t)& \text{and}& & \Gamma_\mathrm{MD}^{(q)}& =F_{\chi_{2}^2\left(\gamma_q\right)}(t),\label{eq:dect}
\end{align}
where $F_{\chi_{2}^2\left(\gamma_q\right)}(t)$ denotes the \gls{cdf} of a $\chi_2^2(\gamma_q)$ distribution.
We observe from \eqref{eq:dect} that for a given desired probability of false alarm, the respective detection threshold $t$ is given by $t=F_{\chi_{2}^2\left(0\right)}^{-1}(1-\Gamma_\mathrm{FA})$.

\section{Phase-Shift Configuration}\label{sec:phasedesign}

The phase-shift vector $\vt{w}$ in \eqref{eq:irsresponse} controls the reflection of the waves impinging on the \gls{irs}. The ideal phase-shift design should provide a low misdetection probability for the entire coverage area because the location $q$ of the active device is not a priori known. Therefore, we formulate a worst-case optimization problem for minimizing the probability of misdetection. Moreover, we propose two heuristic approaches employing closed-form phase-shift vectors.

\subsection{Optimal Phase-Shift Design}

We target a phase-shift design that minimizes the worst-case probability of misdetection $\Gamma_\mathrm{MD}^{(q)}$ across the entire coverage area.
For tractability of the optimization problem, we model the coverage area as a set of $Q$ locations obtained from a grid in the $yz$-plane. Then, every location of interest is index by $q\in\{0, 1, \dots, Q-1\}=\mathcal{Q}$. We note that the grid can be selected to guarantee a desired accuracy, e.g., the grid spacing can be chosen sufficiently small such that two adjacent locations experience approximately the same channel gain.
The optimization objective is the minimization of the largest $\Gamma_\mathrm{MD}^{(q)}$ for $q\in\mathcal{Q}$. However, $\Gamma_\mathrm{MD}^{(q)}$ in \eqref{eq:dect} is a monotonically decreasing function in $\gamma_q$ such that an equivalent objective is the maximization of the smallest $\gamma_q$ for $q\in\mathcal{Q}$.
Moreover, omitting the constant factors in \eqref{eq:gamma_q}, the objective reduces to maximizing the smallest
\begin{equation}\label{eq:objective}
  \abs*{h_q}^2 =\abs*{\frac{\lambda}{4\pi d_r} \frac{\lambda}{4\pi d_t^{(q)}}\upsilon(\vt{\Psi}_t^{(q)}, \vt{\Psi}_r) \vt{a}^H(\vt{\Psi}_t^{(q)}, \vt{\Psi}_r) \vt{w}}^2
\end{equation}
for $q\in\mathcal{Q}$. Using the definitions
\begin{align}
  \bar{\vt{a}}^H(\vt{\Psi}_t^{(q)}, \vt{\Psi}_r)& =\frac{\lambda}{4\pi d_r} \frac{\lambda}{4\pi d_t^{(q)}}\upsilon(\vt{\Psi}_t^{(q)}, \vt{\Psi}_r) \vt{a}^H(\vt{\Psi}_t^{(q)}, \vt{\Psi}_r)\\
  \bar{\mt{A}}(\vt{\Psi}_t^{(q)}, \vt{\Psi}_r)& =\bar{\vt{a}}(\vt{\Psi}_t^{(q)}, \vt{\Psi}_r) \bar{\vt{a}}^H(\vt{\Psi}_t^{(q)}, \vt{\Psi}_r),
\end{align}
we rewrite \eqref{eq:objective} as $\abs*{h_q}^2 = \vt{w}^H \bar{\mt{A}}(\vt{\Psi}_t^{(q)}, \vt{\Psi}_r) \vt{w}$ and formulate the following optimization problem:
\begin{align*}
  &\text{(P1)}&\max_{\vt{w}}\min_{\forall q\in\mathcal{Q}} \quad &\vt{w}^H \bar{\mt{A}}(\vt{\Psi}_t^{(q)}, \vt{\Psi}_r) \vt{w}\\
  &&\text{s.t.} \quad &\abs*{[\vt{w}]_u}=1, \quad \forall\ u\in\mathcal{U}
\end{align*}
Problem (P1) is not convex in $\vt{w}$ due to the unit-modulus constraint \cite{zhang2006complexquadraticoptimization}. A common approach to obtain an approximate solution of (P1) is \gls{sdr} \cite{luo2010semidefiniterelaxationquadratic}.
Using $\mt{W} = \vt{w}\vt{w}^H$ and $\vt{w}^H \bar{\mt{A}}(\vt{\Psi}_t^{(q)}, \vt{\Psi}_r) \vt{w} = \tr{\bar{\mt{A}}(\vt{\Psi}_t^{(q)}, \vt{\Psi}_r) \mt{W}}$,
a relaxed version of (P1) is obtained as
\begin{align*}
  &\text{(P2)}&\max_{\mt{W}, \tau}\quad &\tau\\
  &&\text{s.t.}\quad &\tau\leq \tr{\bar{\mt{A}}(\vt{\Psi}_t^{(q)}, \vt{\Psi}_r)\mt{W}}, \quad \forall\ q\in\mathcal{Q}\\
  &&&\diag(\mt{W}) = \vt{1}\\
  &&&\mt{W} \succeq 0.
\end{align*}
Standard convex solvers, e.g., \cite{mosek2021introduction}, find the optimal solution $\mt{W}_\text{opt}$ of (P2), but $\rank(\mt{W}_\text{opt})=1$ cannot be guaranteed, which means we cannot obtain the optimal phase-shift vector $\vt{w}_\text{opt}$ from $\mt{W}_\text{opt}$ directly.
Instead, we determine an approximation $\hat{\vt{w}}_\text{opt}$ using Gaussian randomization \cite{luo2010semidefiniterelaxationquadratic}:
\begin{enumerate}
  \item For $g\in\{0, 1, \dots, G-1\}=\mathcal{G}$, generate $G$ random vectors $\vt{\nu}_g \sim \mathcal{CN}(0,\mt{W}_\text{opt})$.
  \item For $g\in\mathcal{G}$, set $\abs*{[\vt{\nu}_g]_u}=1\ \forall u\in\mathcal{U}$.
  \item $\hat{\vt{w}}_\text{opt} = \argmax_{\vt{\nu}_g \forall g\in\mathcal{G}} \min_{\forall q\in\mathcal{Q}} \vt{\nu}_g^H \bar{\mt{A}}(\vt{\Psi}_t^{(q)}, \vt{\Psi}_r) \vt{\nu}_g$.
\end{enumerate}
In the remainder of this work, we refer to $\hat{\vt{w}}_\text{opt}$ as the optimized phase-shift design. Problem (P2) is a semi-definite programming problem and, given a solution accuracy $\epsilon>0$, can be solved with a worst-case computational complexity of $\mathcal{O}((U+Q)^4\sqrt{U}\log(1/\epsilon))$ \cite{luo2010semidefiniterelaxationquadratic}. Although there is a polynomial dependency on $U$ and $Q$, the complexity can be afforded because the considered phase-shift design is an offline problem that is solved once in the design stage of the system. Nevertheless, in the next subsection, we propose two closed-form phase-shift designs that entail a lower complexity than the optimized design.

\subsection{Heuristic Phase-Shift Designs}

In the following, we consider analytical phase-shift vector designs characterized by
\begin{equation}\label{eq:phaseshiftentry}
  [\vt{w}]_{u_x,u_y} =
    \exp\left(
      j \begin{bmatrix}u_x & u_y & 0\end{bmatrix}\nabla\omega(u_x, u_y)
    \right),
\end{equation}
where $\nabla\omega(u_x, u_y)\in\mathbb{R}^3$ denotes the phase-shift gradient.

The first design is based on a constant phase gradient that results in a linear phase-shift design. It is known that such a design maximizes the reflection gain for a specific incident direction \cite{najafi2020physicsbasedmodeling}. A straightforward approach chooses the center of the coverage area as the direction for maximum gain, but this leads to poor performance at the corners of the area. Therefore, we propose a design that partitions the \gls{irs} and the coverage area into $K$ tiles, respectively, indexed by $k\in\{0, 1, \dots, K-1\}=\mathcal{K}$. Then, we apply the linear phase-shift design where each tile of the \gls{irs} covers one tile of the coverage area. The partitioning is performed along the $y$-axis and the $k$th tile of the \gls{irs} is specified by $u_y^{(k)}\in\{kU_y/K - U_y/2 + 1, \dots, (k+1)U_y/K - U_y/2\}$.
Furthermore, $\vt{\Psi}_t^{(c_k)}$ denotes the direction from the \gls{irs} to the center of the $k$th tile of the coverage area. This results in the phase gradient
\begin{equation}
  \nabla\omega(u_x, u_y^{(k)}) = -\left(\vt{k}(\vt{\Psi}_t^{(c_k)}) + \vt{k}(\vt{\Psi}_r)\right)
\end{equation}
and specifies the phase-shift vector of the $k$th \gls{irs} tile.

The second design is based on the work in \cite{jamali2021powerefficiencyoverhead}, which uses a linear phase-shift gradient resulting in a quadratic phase-shift design. This approach provides wide coverage by design. The main idea is to determine the required constant phase gradients for every location of the coverage area and obtain their minimum and maximum values. Then, the phase gradient interpolates between these values with a linear function to cover to entire area. To this end, we define the phase gradient
\begin{equation}\label{eq:quadraticphasegradient}
  \nabla\omega(u_x, u_y) =
    \begin{bmatrix}
      u_x & 0 & 0\\
      0 & u_y & 0\\
      0 & 0 & 0
    \end{bmatrix} \begin{bmatrix}\alpha_x\\\alpha_y\\0\end{bmatrix}
    + \begin{bmatrix}\beta_x\\\beta_y\\0\end{bmatrix}
\end{equation}
and determine $\alpha_x$, $\alpha_y$, $\beta_x$, and $\beta_y$ by solving
\begin{equation}\label{eq:quadraticdesign}
  \begin{split}
    \nabla\omega(u_x^\text{min}, u_y^\text{min})& =\min_{\forall q\in\mathcal{Q}} -\left(\vt{k}(\Psi_t^{(q)}) + \vt{k}(\Psi_r)\right)\\
    \nabla\omega(u_x^\text{max}, u_y^\text{max})& =\max_{\forall q\in\mathcal{Q}} -\left(\vt{k}(\Psi_t^{(q)}) + \vt{k}(\Psi_r)\right),
  \end{split}
\end{equation}
where the $\min$ and $\max$ operations are elementwise and $u_x^\text{min}=-U_x/2 + 1$, $u_y^\text{min}=-U_y/2 + 1$, $u_x^\text{max}=U_x/2$, and $u_y^\text{max}=U_y/2$.
Thus, \eqref{eq:quadraticphasegradient} and \eqref{eq:quadraticdesign} specify the phase-shift vector in \eqref{eq:phaseshiftentry}.

\section{Performance Results}\label{sec:results}

\begin{table}
  \scriptsize
  \centering
  \renewcommand{\arraystretch}{1.3}
  \begin{threeparttable}
    \caption{Simulation Parameters}
    \label{tab:parameters}
    \begin{tabular}{l l|l l}
      \hline
      Parameter & Value & Parameter & Value\\
      \hline
      $\lambda$ & \SI{0.1}{\metre} & $(d_x, d_y)$ & $(\lambda /2, \lambda /2)$\\
      $(U_x, U_y)$ & $(8, 8)$ & $(c_x, c_y, c_z)$ & $(\SI{-10}{\metre}, \SI{-50}{\metre}, \SI{50}{\metre})$\\
      $(d_r, \theta_r, \phi_r)$ & $(\SI{30}{\metre}, \SI{0}{\degree}, \SI{90}{\degree})$ & $M$ & 16\\
      $\Gamma_\mathrm{FA}$ & $0.1$ & $P_x$ & \SI{28}{\dBm} \\
      $\sigma_n^2$ & \SI{-95}{\dBm} & $S$ & 32\\
      \hline
    \end{tabular}
  \end{threeparttable}
\end{table}

\begin{figure*}
\centering
\subfloat[Linear Design ($K=1$)]{\includegraphics{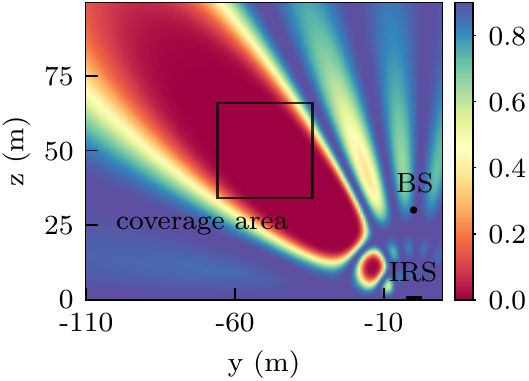}
\label{fig:result1-1}}
\hfil
\subfloat[Quadratic Design]{\includegraphics{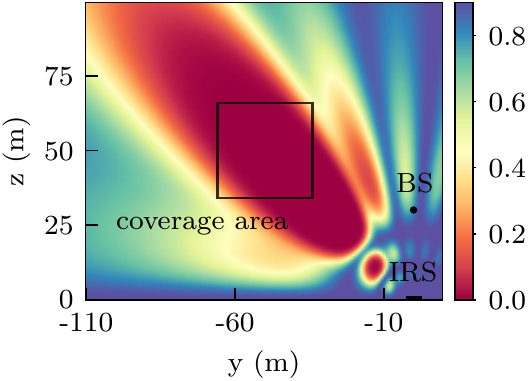}
\label{fig:result1-2}}
\hfil
\subfloat[Optimized Design]{\includegraphics{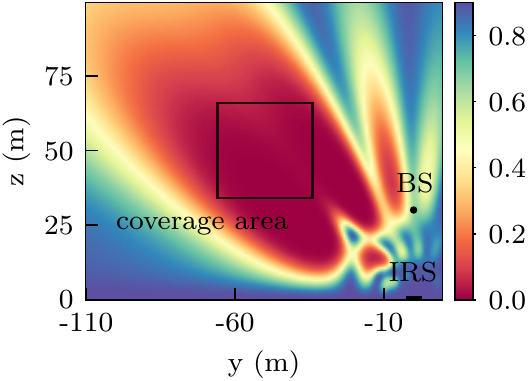}
\label{result1-3}}
\caption{Misdetection probability for different phase-shift designs and area side length $\Delta_y = \Delta_z = \SI{30}{\metre}$.}
\label{fig:result1}
\end{figure*}

In this section, we study the misdetection probability for the proposed phase-shift designs. We set the noise power as $\sigma_n^2=N_0BF=\SI{-95}{\dBm}$ assuming noise \gls{psd} $N_0=\SI{-174}{\dBmpHz}$, signal bandwidth $B=\SI{20}{\mega\hertz}$, and noise figure $F=\SI{6}{\decibel}$. Table \ref{tab:parameters} specifies all relevant system parameters.

To illustrate the impact of different phase-shift designs on the reflection beams of the \gls{irs}, Fig.~\ref{fig:result1} shows the misdetection probability for all points in the $yz$-plane with $x=\SI{-10}{\metre}$, i.e., at the ground. The beam of the linear design with $K=1$ is focused at the center of the coverage area and does not provide sufficient gain in the upper right corner of the area. The quadratic design generates a well aligned beam. However, compared to the optimized design, it is not wide enough to create high gain for the entire coverage area.

\begin{figure}
\centerline{\includegraphics{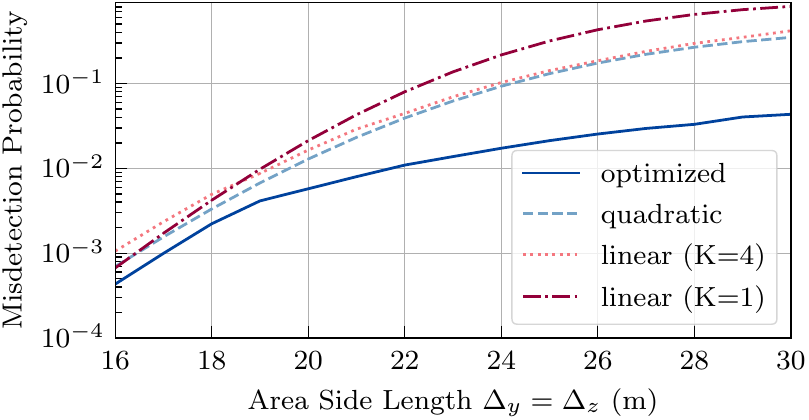}}
\caption{Phase-shift design comparison for \gls{los} channels.}
\label{fig:result2}
\end{figure}
Fig.~\ref{fig:result2} shows the maximum misdetection probability $\Gamma_\mathrm{MD}=\max_{q\in\mathcal{Q}}\Gamma_\mathrm{MD}^{(q)}$ based on \eqref{eq:dect} for the \gls{los} case, i.e., $L=1$.
We plot $\Gamma_\mathrm{MD}$ for the heuristic designs using the analytical phase-shift vector in \eqref{eq:phaseshiftentry} whereas the curve for the optimized design represents the average $\Gamma_\mathrm{MD}$ obtained for $80$ randomized phase-shift vectors $\hat{\vt{w}}_\text{opt}$ with $G=3000$.
We observe that the optimized design provides the best performance. The heuristic designs achieve similar results for small areas, but cannot compete with the optimized design for larger areas. Moreover, for large areas, the linear design with $K=4$ outperforms the design with $K=1$, which confirms that larger areas require a wider reflection beam, i.e., more tiles. This is inherently taken into account by the quadratic design, which yields a better performance than both linear designs for all considered sizes of the coverage area.

\begin{figure}
\centerline{\includegraphics{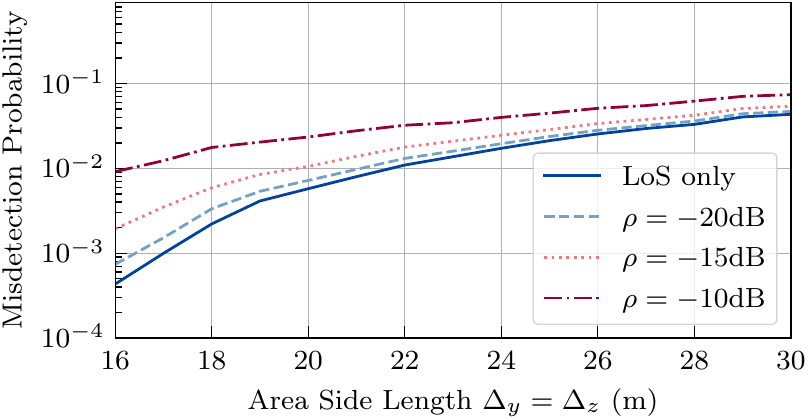}}
\caption{Optimized phase-shift design with scattered device-IRS channel.}
\label{fig:result3}
\end{figure}
The impact of scatterers in the device-\gls{irs} link on the proposed \gls{los}-based design is shown in Fig.~\ref{fig:result3}. We set $L=5$ and $P_{l,\mathrm{NLoS}}^{(q)}=\frac{\rho}{L-1}\abs*{h_{t,0}^{(q)}}^2$, where we vary $\rho$ to evaluate different scattering conditions.
The scatterers' locations are in the local environment of the active device and their incident directions are chosen as $\vt{\Psi}_{t,l>0}^{(q)}\sim\mathcal{N}(\vt{\Psi}_{t,0}^{(q)}, \diag(0.1^2, 0.1^2))$.
Fig.~\ref{fig:result3} shows the average worst-case misdetection probability obtained from Monte-Carlo simulations evaluating \eqref{eq:metric}.
In the presence of scatterers, we observe an increase of $\Gamma_\mathrm{MD}$ compared to the \gls{los} case because the non-\gls{los} components in \eqref{eq:vhp} may add up destructively, leading to fading and a lower received power. The variation of the received power has less impact on $\Gamma_\mathrm{MD}$ when the size of the area is large, as in this case, the size of the coverage area is the performance limiting factor.

\section{Conclusion}\label{sec:conclusion}

This paper studied active device detection in an \gls{irs}-assisted communication system. We derived a \gls{glrt} detector and an optimized worst-case phase-shift design for a given coverage area. Besides, we proposed two heuristic phase-shift designs. Our performance comparison showed the superiority of the optimized design and demonstrated the impact of scattering on the \gls{los}-based designs. More sophisticated phase-shift designs that take into account the impact of scattering constitute an interesting topic for future work.

\bibliographystyle{IEEEtran}
\bibliography{ieeetrancontrol,IEEEabrv,references}
\end{document}